\documentstyle[12pt]{article}

\def\hybrid{\topmargin -20pt	\oddsidemargin 0pt
	\headheight 0pt	\headsep 0pt
	\textwidth 6.25in	
	\textheight 9.5in	
	\marginparwidth .875in
	\parskip 5pt plus 1pt	\jot = 1.5ex}

\hybrid

\def\baselinestretch{1.2}

\catcode`\@=11

\def\marginnote#1{}
\newcount\hour
\newcount\minute
\newtoks\amorpm
\hour=\time\divide\hour by60
\minute=\time{\multiply\hour by60 \global\advance\minute by-\hour}
\edef\standardtime{{\ifnum\hour<12 \global\amorpm={am}%
	\else\global\amorpm={pm}\advance\hour by-12 \fi
	\ifnum\hour=0 \hour=12 \fi
	\number\hour:\ifnum\minute<10 0\fi\number\minute\the\amorpm}}
\edef\militarytime{\number\hour:\ifnum\minute<10 0\fi\number\minute}

\def\draftlabel#1{{\@bsphack\if@filesw {\let\thepage\relax
   \xdef\@gtempa{\write\@auxout{\string
      \newlabel{#1}{{\@currentlabel}{\thepage}}}}}\@gtempa
   \if@nobreak \ifvmode\nobreak\fi\fi\fi\@esphack}
	\gdef\@eqnlabel{#1}}
\def\@eqnlabel{}
\def\@vacuum{}
\def\draftmarginnote#1{\marginpar{\raggedright\scriptsize\tt#1}}

\def\draft{\oddsidemargin -.5truein
	\def\@oddfoot{\sl preliminary draft \hfil
	\rm\thepage\hfil\sl\today\quad\militarytime}
	\let\@evenfoot\@oddfoot	\overfullrule 3pt
	\let\label=\draftlabel
	\let\marginnote=\draftmarginnote
   \def\@eqnnum{(\theequation)\rlap{\kern\marginparsep\tt\@eqnlabel}%
\global\let\@eqnlabel\@vacuum}  }

\def\preprint{\twocolumn\sloppy\flushbottom\parindent 2em
	\leftmargini 2em\leftmarginv .5em\leftmarginvi .5em
	\oddsidemargin -.5in	\evensidemargin -.5in
	\columnsep .4in	\footheight 0pt
	\textwidth 10.in	\topmargin  -.4in
	\headheight 12pt \topskip .4in
	\textheight 6.9in \footskip 0pt
	\def\@oddhead{\thepage\hfil\addtocounter{page}{1}\thepage}
	\let\@evenhead\@oddhead	\def\@oddfoot{}	\def\@evenfoot{} }

\def\numberbysection{\@addtoreset{equation}{section}
	\def\theequation{\thesection.\arabic{equation}}}

\def\underline#1{\relax\ifmmode\@@underline#1\else
	$\@@underline{\hbox{#1}}$\relax\fi}

\def\titlepage{\@restonecolfalse\if@twocolumn\@restonecoltrue\onecolumn
     \else \newpage \fi \thispagestyle{empty}\c@page\z@
	\def\thefootnote{\fnsymbol{footnote}} }

\def\endtitlepage{\if@restonecol\twocolumn \else \newpage \fi
	\def\thefootnote{\arabic{footnote}}
	\setcounter{footnote}{0}}  

\catcode`@=12
\relax

\def\figcap{\section*{Figure Captions\markboth
	{FIGURECAPTIONS}{FIGURECAPTIONS}}\list
	{Figure \arabic{enumi}:\hfill}{\settowidth\labelwidth{Figure
999:}
	\leftmargin\labelwidth
	\advance\leftmargin\labelsep\usecounter{enumi}}}
 \relax
\def\tablecap{\section*{Table Captions\markboth
	{TABLECAPTIONS}{TABLECAPTIONS}}\list
	{Table \arabic{enumi}:\hfill}{\settowidth\labelwidth{Table
999:}
	\leftmargin\labelwidth
	\advance\leftmargin\labelsep\usecounter{enumi}}}
 \relax
\def\reflist{\section*{References\markboth
	{REFLIST}{REFLIST}}\list
	{[\arabic{enumi}]\hfill}{\settowidth\labelwidth{[999]}
	\leftmargin\labelwidth
	\advance\leftmargin\labelsep\usecounter{enumi}}}
 \relax
\makeatletter
\newcounter{pubctr}
\def\publist{\@ifnextchar[{\@publist}{\@@publist}}
\def\@publist[#1]{\list
	{[\arabic{pubctr}]\hfill}{\settowidth\labelwidth{[999]}
	\leftmargin\labelwidth
	\advance\leftmargin\labelsep
	\@nmbrlisttrue\def\@listctr{pubctr}
	\setcounter{pubctr}{#1}\addtocounter{pubctr}{-1}}}
\def\@@publist{\list
	{[\arabic{pubctr}]\hfill}{\settowidth\labelwidth{[999]}
	\leftmargin\labelwidth
	\advance\leftmargin\labelsep
	\@nmbrlisttrue\def\@listctr{pubctr}}}
 \relax
\makeatother
\newskip\humongous \humongous=0pt plus 1000pt minus 1000pt

\newif\ifdtup

\relax

\def\thefootnote{\fnsymbol{footnote}}
\def\be{\begin{equation}}
\def\ee{\end{equation}}
\def\ba{\begin{eqnarray}}
\def\ea{\end{eqnarray}}

\begin{document}
\renewcommand{\theequation}{\arabic{equation}}
\newcommand{\beq}{\begin{equation}}
\newcommand{\eeq}[1]{\label{#1}\end{equation}}
\newcommand{\ber}{\begin{eqnarray}}
\newcommand{\eer}[1]{\label{#1}\end{eqnarray}}
\begin{titlepage}
\begin{center}

\hfill CERN--TH.7473/94\\
\hfill hep-th/9410104\\

\vskip .4in

{\large \bf SPACE TIME INTERPRETATION OF S--DUALITY AND\\
            SUPERSYMMETRY VIOLATIONS OF T--DUALITY}

\vskip .7in

{\bf Ioannis Bakas}
\footnote{Permanent address: Department of Physics, University of
Crete,
GR--71409 Heraklion, Greece}
\footnote{e--mail address: BAKAS@SURYA11.CERN.CH}
\vskip .2in

{\em Theory Division, CERN\\
     CH-1211 Geneva 23, Switzerland}\\

\vskip .1in

\end{center}

\vskip .5in

\begin{center} {\bf ABSTRACT } \end{center}
\begin{quotation}\noindent
The S--duality transformations of the lowest order string effective
theory admit a space time interpretation for 4--dim
backgrounds with one Killing symmetry. Starting from pure gravity
and performing a sequence of intertwined T--S--T duality
transformations we obtain new solutions which are always pure
gravitational. In this fashion, S--duality induces an $SL(2,R)$
transformation in the space of target space metrics which coincides
with the action of the Ehlers--Geroch group and interchanges the
electric with the magnetic aspects of gravity. Specializing to
gravitational instanton backgrounds we show that ALE instantons
are mapped to (multi) Taub--NUT backgrounds and vice--versa.
We find, however, that the self--duality of the metric is not
generically preserved, unless the corresponding Killing vector
field has self--dual covariant derivatives. Thus, the T--S--T
transformations are not always compatible with the world--sheet
supersymmetry of $N=4$ superconformal string vacua. We also provide
an algebraic characterization of the corresponding obstruction
and associate it with a breakdown of space time supersymmetry
under rotational T--duality transformations.

\end{quotation}
\vskip.5cm
CERN--TH.7473/94 \\
October 1994\\
\end{titlepage}
\vfill
\eject

\def\baselinestretch{1.2}
\baselineskip 16 pt

The toroidal compactification of the heterotic string to $M_{4}
\times T^{6}$
seems to exhibit a remarkable symmetry under an $SL(2, Z)$ group of
transformations which act on the coupling constant of the theory [1,
2]. This
discrete symmetry, known as S--duality, has its origin in the lowest
order 4--dim effective theory where the axion--dilaton system admits
a
continuous $SL(2, R)$ symmetry that becomes manifest in the Einstein
frame.
The purpose of this paper is to show that S--duality can have a
space time interpretation when it is appropriately combined with
T--duality associated to a Killing symmetry on $M_{4}$.
If both symmetries are valid and remain consistent with
supersymmetry,
then different space time backgrounds of the heterotic string will be
related. We will see, for instance, that in the toroidal
compactification
of the heterotic string, S--duality relates vacua with $M_{4}$ being
either the flat or the Taub--NUT space for certain discrete values of
its
moduli parameter. It is not surprising that both these spaces admit
four
standard space time supersymmetries [3].

We consider bosonic backgrounds which are supersymmetric solutions of
the heterotic string and so all fermion fields together with their
supersymmetric variations are taken zero. As for the gauge fields,
the
standard embedding is implicitly used, in order to insure that all
higher
order ${\alpha}^{\prime}$ corrections which are proportional to
$Tr R \wedge R - Tr F \wedge F$ vanish. The known bosonic solutions
of the
heterotic string correspond to supersymmetric self--dual backgrounds
for
which the standard embedding of the gauge fields $A$ amounts to the
self--duality condition on $F(A)$ [4--6]. We use this as a strarting
point to show
that if a self--dual background admits Killing symmetries, its
T--dual will
not necessarily be supersymmetric and hence it will not always
qualify for
a new bosonic solution of the heterotic superstring. As we will see
later, there are two types of Killing vector fields distinguished by
the
requirement of preserving standard supersymmetry or not. The first
type,
called translational, has self--dual covariant derivatives and
preserves
the vanishing condition of the supersymmetric variations of the
fermion
fields. There exist, however, rotational Killing vector fields, in
that
their covariant derivatives are not self--dual, which are not
compatible with space time supersymmetry. The ordinary gravitational
instantons (including flat space) always admit at least one
translational
Killing symmetry and so their T--dual versions are legitimate
solutions.
But if one chooses a rotational Killing symmetry, which is present in
simple examples like the flat space and the Eguchi--Hanson instanton,
then
a classical supersymmetric anomaly appers by performing the
corresponding T--duality transformation.
This result casts doubts on the string equivalence of T--duality
related backgrounds and calls for a careful investigation in the
general
supersymmetric case [7].

Most of the present work concentrates on the properties and
symmetries of the
lowest order 4--dim effective bosonic theory [8],
\be
S_{eff} = \int_{{M}_{4}} d^{4}X \sqrt{\det G}~ e^{-2 \Phi}
\left(R[G]
+4 {(\nabla \Phi)}^{2} - {1 \over 12} H^{2} \right),
\ee
after supressing the contribution of the gauge fields.
Here $H_{\mu \nu \rho} = 3 \nabla_{[ \mu} B_{\nu \rho]}$ is the field
strength
of the anti--symmetric tensor field $B_{\mu \nu}$ and $\Phi$ is the
dilaton
field. The cosmological constant (central charge deficit) is also
taken to be
zero in the lowest order effective action. From now on, we will
consider the general class of 4--dim metrics with (at least) one
Killing
symmetry associated to a vector $K = \partial / \partial \tau$ on
$M_{4}$. Then, it
is well known that any such metric can be written locally in the form
\begin{equation}
{d s}^{2} = V {(d \tau + {\omega}_{i} d x^{i})}^{2} +
V^{-1} {\gamma}_{ij} d x^{i} d x^{j} ~ ,
\end{equation}
where $\{ x^{i}~ ; ~ i = 1, 2, 3 \}$ are coordinates on the space of
non--trivial orbits of
$ \partial / \partial \tau$ in $M_{4}$ and $V,~ {\omega}_{i},~
{\gamma}_{ij}$ are
all independent of $\tau$, but otherwise arbitrary. The results we
will describe
are quite general, but some of their implications will be considered
in the
context of the heterotic string theory.

The solutions of the vacuum Einstein equations can be regarded as
special
cases of gravitational string backgrounds with zero dilaton $\Phi$
and
anti--symmetric tensor field $B_{\mu \nu}$.
In view of this embedding, if we perform a T--duality transformation
(see for
instance [9] and references therein)
\ba
{\tilde{G}}_{\tau \tau} & = & {1 \over G_{\tau \tau}} ~ , ~~~
{\tilde{G}}_{\tau i} = {B_{\tau i} \over G_{\tau \tau}} ~, ~~~
{\tilde{G}}_{ij} = G_{ij} - {G_{\tau i} G_{\tau j} - B_{\tau i}
B_{\tau j}
\over G_{\tau \tau}} ~,\\
{\tilde{B}}_{\tau i} & = & {G_{\tau i} \over G_{\tau \tau}} ~, ~~~
{\tilde{B}}_{ij} = B_{ij} - {G_{\tau i} B_{\tau j} - G_{\tau j}
B_{\tau i}
\over G_{\tau \tau}} ~,\\
{\tilde{\Phi}} & = & \Phi - {1 \over 2} \log G_{\tau \tau}
\ea
to the pure
gravitational metric (2),
new solutions to the string background equations
will result with non--trivial $\Phi$ and $B_{\mu \nu}$ fields in
general.

On the other hand, the effective theory (1) exhibits an additional
global
$SL(2, R)$ symmetry which is manifestly described in
terms of the axion--dilaton system
formulated in the Einstein frame of the string,
\be
G^{(E)}_{\mu \nu} = e^{-2 \Phi} G_{\mu \nu} ~.
\ee
In this frame, the axion field $b$ can be consistently defined as
\be
{\partial}_{\mu} b = {e^{-4 \Phi} \over 6} \sqrt{\det G^{(E)}}
{{\epsilon}_{\mu}}^{\nu \rho \sigma} H_{\nu \rho \sigma} ~,
\ee
where ${\epsilon}_{\tau 1 2 3} = 1$, and so $b$ is non--locally
related to the
original variables. The axion--dilaton system behaves
in the Einstein frame as an $SL(2, R)/U(1)$
non--linear $\sigma$--model,
which in terms of the pair of conjugate variables
\be
S_{\pm} = b \pm e^{-2 \Phi}
\ee
remains invariant under the $SL(2, R)$ group of transformations
\be
S_{\pm} \rightarrow {A S_{\pm} + B \over C S_{\pm} + D} ~, ~~~ AD-BC
= 1.
\ee
We will refer to this symmetry of the 4--dim effective theory as
S--duality, although it is appropriate to reserve the name
duality for discrete string symmetries and not just for continuous
symmetries of the lowest order effective theory. The
$SL(2, Z)$ string duality symmetry will be considered later, when
additional restrictions will be also imposed on the background
metric.

The main observation is that by performing a sequence of T--S--T
duality tranformations
on a pure gravitational background with one Killing symmetry,
the resulting string background is pure gravitational as well. Since
the
loop of T--S--T operations turns out to be
a generic $SL(2, R)$ transformation in the
space of target space metrics, it can be regarded as providing a
space time
interpretation of S--duality. Moreover, as we are going to show, this
$SL(2, R)$ transformation in the space of Ricci flat metrics
coincides with the action of the Ehlers--Geroch symmetry in ordinary
general relativity upon reduction from four to three dimensions [10,
11].
Therefore, thinking of ordinary gravity as a special case of the
4--dim
string background equations, we can provide an alternative
description
of the Ehlers--Geroch symmetry as being induced by S--duality
by employing suitably T--duality on the way, switching on and off
non--trivial
axion and dilaton fields. Another advantage of this
interpretation that will be discussed later, is that S--duality
can be consequently understood as a symmetry interchanging the
electric
with the magnetic
aspects of gravity, since to each metric of the form (2) there is
associated a natural Maxwell field,
\be
A = V (d \tau + {\omega}_{i} d x^{i}) ~.
\ee

We proceed by performing first the T--duality transformation (3)--(5)
to the metric (2)
within the string effective theory (1). The result of the calculation
in the
Einstein frame of the string can be summarized as follows,
\be
{\tilde{G}}^{(E)}_{\mu \nu} = \left( \begin{array}{cccc}
              1& 0 & 0 & 0\\
              0&   &   &  \\
              0&   & {\gamma}_{ij} & \\
              0&   &     & \end{array} \right) ~
\ee
and
\be
{\tilde{\Phi}} = - {1 \over 2} \log V ~, ~~~ {\tilde{B}}_{\tau i} =
{\omega}_{i} ~,
\ee
while all other components are zero. Of course ${\omega}_{i}$ (and
hence
${\tilde{B}}_{\tau i}$) are defined up to a gauge transformation
\be
{\omega}_{i} \rightarrow {\omega}_{i} - {\partial \lambda \over
\partial x^{i}}~,
\ee
which amounts to the coordinate (shift) transformation
\be
\tau \rightarrow \tau + \lambda(x^{i}) ~.
\ee
Using the defining relations for the axion field $b$, we find in this
case
the result
\be
{\partial}_{i} b = {1 \over 2} V^{2} \sqrt{\det \gamma} ~
{{\epsilon}_{i}}^{jk}
({\partial}_{j} {\omega}_{k} - {\partial}_{k} {\omega}_{j}) ~,
\ee
since ${\epsilon}_{\tau ijk} = {\epsilon}_{ijk}$ in the metric (11).
In the terminology of Gibbons and Hawking [11], $b$ coincides with
the so called
nut potential of the metric (2). The nut charge of a gravitational
solution
can be interpreted as the axion charge of the T--dual string
solution.
To perform the S--duality on the resulting
axion--dilaton system we consider the corresponding pair of conjugate
fields
\be
S_{\pm} = b \pm e^{-2 \Phi} = b \pm V ~.
\ee

It can be easily verified that the $SL(2, R)$ transformation (9)
reads for the $V$ and $b$ fields as
\be
V^{\prime} = {V \over (Cb + D)^{2} - C^{2}V^{2}} ~,
\ee
\be
b^{\prime} = {(AD + BC)b + AC(b^{2} - V^{2}) + BD \over
(Cb + D)^{2} - C^{2}V^{2}} ~,
\ee
while the 3--metric ${\gamma}_{ij}$ remains invariant. After this
transformation,
the new axion field can be written again in the form (15) using the
resulting expression (17) for the new $V^{\prime}$ and assuming the
existence of
a new vector field ${\omega}^{\prime}_{i}$. Hence, applying again a
T--duality
transformation, it is obvious that closing the loop of T--S--T
operations yields
a pure gravitational background of the form (2) for the primed
variables
$V^{\prime}$ and ${\omega}^{\prime}_{i}$, but with the same
${\gamma}_{ij}$.
The computation shows that ${\omega}^{\prime}_{i}$ is only implicitly
determined by $V$ and the nut potential $b$ of the original metric in
terms of
the coupled system of equations
\be
{1 \over 2} \sqrt{\det \gamma}~ {{\epsilon}_{i}}^{jk}
({\partial}_{j}{\omega}^{\prime}_{k} -
{\partial}_{k}{\omega}^{\prime}_{j}) =
{1 \over V^{2}} \left( \left((Cb + D)^{2} + C^{2}V^{2}\right)
{\partial}_{i}b
- 2CV (Cb + D) {\partial}_{i}V \right) ~.
\ee
Clearly, ${\omega}^{\prime}_{i}$
are non--local expressions of the original data, but it is important
for our
purposes that they exist and they are unique, up to gauge
transformations
of the form (13).

The result we have just described, obvious as it may seem in
retrospect,
provides a space time interpretation of S--duality, which in the
present
context coincides with the Ehlers--Geroch $SL(2, R)$ symmetry group
of
vacuum Einstein spaces with (at least) one Killing symmetry. To make
the
identification exact, note that
group elements of the form $C=0, ~ A=D=1$ shift $b$ by a constant $B$
and keep
$V$ invariant (translations), while group elements of
the form $B = C = 0, ~ D = A^{-1}$ scale
both $V$ and $b$ by a constant $A^{2}$ (dilations). The really
non--trivial part
of the symmetry is described by the group element with $A=D=1$ and
$B=0$, which
is nothing else but the celebrated Ehlers transform of the 4--dim
metric.
It is also interesting to note that from this point of view, the
three
parameter group of transformations (9) describes a duality between
the
electric aspects of gravity, characterized by $V$, and the magnetic
aspects
characterized by the nut potential $b$ (being the axion field in the
T--dual string setting
of the problem). These electric and magnetic aspects of gravity can
be best
understood by computing the components of the curvature of the
Maxwell field (10)
associated with any metric of the form (2). Hence, S--duality can be
regarded
once more as being analogous to the well known duality symmetry of
electromagnetism.

We are considering 4--dim space times with one Killing symmetry and
so
the isometry group that is generated by the vector field
$\partial / \partial \tau$ might have fixed points
in its action. It is known that the corresponding fixed points are
either
isolated (called nuts) or they form 2-dim submanifolds (called
bolts). Bolts are
thought to be analogous to electric type mass--monopoles, while nuts
as
gravitational dyons with real electric mass--monopole and imaginary
magnetic
mass--monopole [11]. The existence of nuts and bolts in Euclidean
gravity and the relation
between them is a manifestation of the space time interpretation of
S--duality we have presented. Single nuts are not physically
acceptable
in classical relativity, but
bolts are, because of the asymptotic behaviour of the corresponding
solutions.
However, if a string theory had $SL(2, Z)$ S--duality as an exact
symmetry, then
roughly speaking there would be an equivalence
mapping between nuts and bolts in that
string framework. We will make this idea concrete next in the context
of
$N = 4$ superconformal string theory, where S--duality can
relate asymptotically locally
Euclidean (ALE) gravitational instanton
backgrounds to (multi) Taub--Nut backgrounds,
and vice--versa.

It is an established result in the literature that for pure
gravitational
backgrounds, $N=4$ supersymmetry on the world--sheet implies the
hyper--Kahler
condition on the metric. In four dimensions this means that Ricci
flat
metrics are self--dual (or anti self--dual), i.e.,
\be
R_{\mu \nu \rho \sigma} = \pm {1 \over 2} \sqrt{\det G} ~
{{\epsilon}_{\rho \sigma}}^{\kappa \lambda} R_{\mu \nu \kappa
\lambda} ~
\ee
and so they qualify as gravitational instanton backgrounds.
We will consider the self--dual case picking up the plus sign, while
the
anti self--dual case can be treated in a similar way. One expects
that if
T--duality always preserves space time supersymmetry in $N=4$
superconformal
string theory,
the application of a T--S--T sequence of operations
on self--dual metrics with one Killing symmetry will preserve
self--duality
and provide an S--duality relation between gravitational instanton
backgrounds.
In the heterotic string theory it is sufficient to consider the
lowest order effective
action, since all higher order corrections in ${\alpha}^{\prime}$ can
be
made zero by the standard embedding of the gauge fields.
Also, it is sufficient for our argument to consider only the
bosonic sector of the theory and reexamine the action of the
Ehlers--Geroch
group on self--dual metrics. It will turn out that some surprises
arise for
a certain class of metrics satisfying various additional technical
conditions,
while testing the T--S--T duality transformations.
We will also show that for self--dual backgrounds
of ALE or Taub--Nut type the self--duality condition can be preserved
by the transformation, but again not always.

Self--dual metrics of the form (2) can be classified into two classes
depending
on the type of the Killing symmetry. In particular, there are two
possibilities
depending on whether the covariant derivative of the Killing vector
field
$K_{\nu}$, ${\nabla}_{\mu} K_{\nu}$, is self--dual or not [12].
The algebraic condition that leads to this
distinction is solely described by the characteristic quantity
\be
\Delta S_{-} \equiv {\gamma}^{ij} {\partial}_{i} S_{-} {\partial}_{j}
S_{-} ~,
\ee
where $S_{-}$ is one of the two conjugate variables (8) involving the
nut
potential $b$ (equivalently the axion in the T--dual formulation) and
$V$ of
the original metric.
Since the 3--metric ${\gamma}_{ij}$ has Euclidean signature $+++$,
the
quantity $\Delta S_{-}$ is non--negative. The two different cases in
question
arise when $\Delta S_{-} = 0$ or $\Delta S_{-} > 0$.\footnote{If
we were considering anti self--dual metrics
then there would be an analogous condition on $S_{+}$ instead, which
in more
geometrical terms would be equivalent to having an anti self--dual
${\nabla}_{\mu} K_{\nu}$ or not. The reason is
that if a metric of the form (2) is self--dual, its anti self--dual
counterpart will be obtained by flipping the sign of all
${\omega}_{i}$
and hence of the corresponding $b$. Under this interchange we clearly
have
$S_{\pm} \rightarrow -S_{\mp}$.}
In the first case the
Killing symmetry is usually called ``translational", while in the
second case
``rotational", although the real distiction between the two depends
on the
quantity (21) being zero or strictly positive. Of course, it might be
possible for a self--dual metric to have more than one Killing
symmetry with
respect to which is either of one or of the other type; the
Eguchi--Hanson
instanton is the simplest non--trivial example with two such
different types of
Killing symmetries, but the flat space can entertain both
possibilities as well.
It is important to realize that the analysis we will
present in the sequel depends crucially on the Killing symmetry that
is
chosen to perform the T--S--T transformation.

In the first case, when $\Delta S_{-} = 0$ and hence $S_{-}$ is
constant,
there is a theorem [12, 13] stating that there exists a coordinate
system in which
${\gamma}_{ij} = {\delta}_{ij}$ and the self--duality condition
amounts to
the condition
\be
{\partial}_{i} V^{-1} = {1 \over 2} {\epsilon}_{ijk}
({\partial}_{j}{\omega}_{k}
- {\partial}_{k}{\omega}_{j}) ~,
\ee
with respect to the flat 3--metric ${\delta}_{ij}$. Hence, $V^{-1}$
satisfies
the 3--dim Laplace equation and all components of the 4--dim metric
are
determined in terms of its solutions. In this case, comparing eqs.
(15) and
(22), we find that
\be
S_{+} = 2V ~, ~~~~ S_{-} = 0 ~,
\ee
up to an overall constant which is taken to be zero with no loss of
generality.
Solutions to the 4--dim string background equations, where a
(conformally) flat
metric is coupled to axionic instantons have been considered before
[4, 5], together
with their T--dual formulations as pure gravitational instanton
backgrounds [6].
The main issue  of our concern is not to rederive this result, but to
note that
the sequence of operations T--S--T preserves self--duality and the
characteristic condition $\Delta S_{-} = 0$. Since the only
non--trivial part
of the Ehlers--Geroch symmetry corresponds to the choice $B=0, ~
A=D=1$,
applying formula (17) we find the effect of the Ehlers transformation
to be
\be
V^{-1} \rightarrow V^{-1} + 2C ~,
\ee
where in the discrete $SL(2, Z)$ case $C$ is an arbitrary integer.
Its effect on ${\omega}_{i}$ is trivial and up to gauge
transformations we may
take ${\omega}^{\prime}_{i} = {\omega}_{i}$.  This shift
is clearly a symmetry of the 3--dim Laplacian equation imposed on
$V^{-1}$
by the self--duality and
although it is rather simple, it can have important effects on the
space time
interpretation of the corresponding solutions.

The localized solutions of the 3--dim
Laplace equation for $V^{-1}$ are of the general form
\be
V^{-1} = \epsilon + \sum_{i=1}^{n} {m_{i} \over \mid \vec{x} -
{\vec{x}}_{0,i} \mid} ~,
\ee
modulo delta functions,
where $m_{i}$ and ${\vec{x}}_{0,i}$ are moduli parameters. The
constant $\epsilon$
can be either zero or non--zero, in which case it is usually
normalized to 1.
If all $m_{i} = M$ and if $\tau$ is periodic with the range
$0 \leq \tau \leq 4 \pi M / n$, then it is known that the
singularities of the
corresponding 4--dim self--dual metric are removable and the
solutions could
qualify as gravitational instanton backgrounds [11] (see also [14]
for a review).
{}From now on, the singularities
of the metric, being either nuts or bolts depending on the coordinate
system,
will be assumed to be removable. For $\epsilon = 1$ one has the
multi  Taub--NUT metrics, with $n=1$ being the ordinary self--dual
Taub--NUT
metric. For $\epsilon = 0$ the resulting metrics are the
multi--center
Gibbons--Hawking metrics, with $n=2$ being the simplest non--trivial
example
known as the Eguchi--Hanson instanton. They correspond to the
A--series in the
A--D--E classification of the ALE gravitational instantons [15].
For $\epsilon = 0$ and $n=1$ one obtains
the flat metric space by appropriate coordinate transformation. In
gravitational
theories, only the solutions with
$\epsilon = 0$ are physically acceptable based on their
asymptotically locally Euclidean behaviour; the boundary at infinity
is the lens
space $S^{3} / Z_{n}$. For the multi Taub--NUT metrics, the existence
of a non--zero
nut parameter violates the ALE asymptotic condition and the solutions
are
asymptotically flat only in the spatial direction
$\mid \vec{x} \mid \rightarrow \infty$, but they are periodic in the
variable
$\tau$. In other words, $\epsilon$ being zero or not changes the
sense in which
the solution is asymptotically flat from the four to the three
dimensional case.

It is often the case with symmetry groups of differential equations
that their
action on the space of solutions does not respect boundary
conditions. In the
space time interpretation we are presently considering, this means
that
S--duality can in principle relate geometrically (and even
topologically)
different backgrounds.\footnote{The string equivalence between
geometrically
and even topologically different backgrounds which are related by
T--duality
was considered before [16].} In particular, under the Ehlers
transform
ALE instantons are mapped to multi Taub--NUT backgrounds and vice
versa, as
it was already indicated.
This changes the geometry and the asymptotic behaviour of the
corresponding solutions. Suppose that we start from a multi--center
Gibbons--Hawking instanton and set $M=1$ by rescaling simoultaneously
all the space time coordinates. Then, the Ehlers transform of
$V^{-1}$
gives rise to a multi Taub--NUT space with $M = 1 / 2C$, provided
that
$C \neq 0$. The period of $\tau$ scales accordingly under this
transformation.
Reversely, starting from  multi Taub--NUT backgrounds it is
possible to obtain the
multi--center Gibbons--Hawking instantons. This is more easily seen
if
$\epsilon$ is not normalized to 1 but it is allowed to assume
arbitrary
non--zero values; otherwise a composition of the Ehlers transform
(24)
with the dilations will be necessary to complete the mapping.
It is appropriate to consider $C > 0$ in all these cases.

The Ehlers transform maps the flat space to the ordinary Taub--NUT
space.
This case corresponds to self--dual metrics with a single center
which
can be positioned at the origin using the freedom of translations in
$\vec{x}$--space. The metric with $\epsilon = 0$ has $V = \mid
\vec{x} \mid$
and
\be
{\omega}_{1}  =  - {x^{2} x^{3} \over \mid \vec{x} \mid
\left( {(x^{1})}^{2} + {(x^{2})}^{2} \right)} ~ , ~~~
{\omega}_{2} =  {x^{1} x^{3} \over \mid \vec{x} \mid
\left( {(x^{1})}^{2} + {(x^{2})}^{2} \right)} ~ , ~~~
{\omega}_{3} = 0 ~ .
\ee
If we introduce two complex coordinates $z_{1}$ and $z_{2}$ so that
\ba
x_{1} & = & {1 \over 2} Im ({\bar{z}}_{1} z_{2}) ~ , ~~~~
x_{2} = {1 \over 2} Re ({\bar{z}}_{1} z_{2}) ~ ,\nonumber\\
x_{3} & = & {1 \over 4} \left( {\mid z_{1} \mid}^{2} -
{\mid z_{2} \mid}^{2} \right) ~ , ~~~~
\tau = Im \log (z_{1} z_{2}) ~ ,
\ea
the metric assumes the flat form with $z_{1}$ and $z_{2}$ as its
corresponding
Kahler coordinates. Its Ehlers transform with respect to the Killing
vector
field $\partial / \partial \tau$ is the Taub--NUT metric. The period
of the
$\tau$ coordinate is $2 \pi / C$ after removing the singularity.
Since both
these 4--dim backgrounds admit $N=4$ space time supersymmetry [3], it
is
natural to expect that in the heterotic string theory they will be
related by S--duality for discrete values of $C$. This novel
possibility has
to be tested directly. It is curious to note that the compactified
Schwarzschild solution can be regarded as a combination of a
self--dual Taub--NUT
metric with its anti self--dual partner, in order to cancel the
magnetic mass.

We will now examine the other general class of self--dual metrics
with one
rotational symmetry corresponding to the algebraic condition $\Delta
S_{-} > 0$.
There is a theorem [13] stating that in this case the self--dual
metrics
are determined by a single scalar function $\Psi(x^{i})$ in a
coordinate
system of the form (2) so that
\be
{\gamma}_{11} = {\gamma}_{22} = e^{\Psi} ~, ~~~~ {\gamma}_{33} = 1 ~,
\ee
while the off--diagonal components of the 3--metric are zero and
\be
V^{-1} = {\partial}_{3} \Psi ~, ~~~ {\omega}_{1} = - ~ {\partial}_{2}
\Psi ~,
{}~~ {\omega}_{2} = {\partial}_{1} \Psi ~, ~~~ {\omega}_{3} = 0 ~.
\ee
Then, in terms of this parametrization the self--duality condition
for the
4--metric becomes the continual Toda equation for $\Psi$, namely
\be
({{\partial}_{1}}^{2} + {{\partial}_{2}}^{2}) \Psi +
{{\partial}_{3}}^{2}
e^{\Psi} = 0 ~.
\ee
Performing the T--duality on such a gravitational background we find
that
\be
S_{+} = 2V - x^{3} ~, ~~~~ S_{-} = - ~ x^{3} ~,
\ee
up to an overall constant and with Einstein metric given by eq. (11).
In this coordinate system $\Delta S_{-}$ remains positive (as it
should be
for a coordinate independent description of the characteristic
condition)
and takes the value 1.

It is interesting to note at this point that the space of solutions
of the
continual Toda equation exhibits infinitely many symmetries
generating a
classical $W_{\infty}$ algebra [17]. The simplest representative
(associated with
the centerless Virasoro algebra) is given infinitesimally by the
transformation
\be
\delta \Psi = \partial \epsilon + \epsilon ~ \partial \Psi ~,
\ee
where the infinitesimal parameter $\epsilon$ depends on the chiral
combination
$2z = x^{1} + i x^{2}$ and the derivatives are taken with respect to
$z$.
There is a similar transformation for the other chiral sector in
terms of
$2\bar{z} = x^{1} - i x^{2}$. Higher spin symmetries have been also
studied,
but their form is quite complicated. The continual Toda equation also
exhibits
a global $U(1)$ symmetry
\be
\delta \Psi = \epsilon ~ {\partial}_{3} \Psi ~,
\ee
where $\epsilon$ is now taken to be independent of all space time
variables.
These symmetries, by construction, transform any self--dual
gravitational
background with $\Delta S_{-} > 0$ into another, while preserving
self--duality.
They can certainly be employed to induce a mapping between
non--trivial
string backgrounds by intertwining them with T--duality. From this
point of
view one might suspect that the S--duality, if it always
preserves the self--duality of
the original metric, will be identified with a subgroup of the
infinite
dimensional symmetry group of the continual Toda theory. However, as
we will
see shortly this turns out not to be the case and we discover an
anomaly for
$\Delta S_{-} \neq 0$.

The simplest set of solutions of the continual Toda field equation
can be
obtained by making the ansatz
\be
e^{\Psi(x^{1}, x^{2}, x^{3})} = \left( \alpha {(x^{3})}^{2} + \beta
x^{3}
+ \gamma \right) e^{\varphi(x^{1}, x^{2})} ~.
\ee
It is straightforward to verify that this ansatz will provide
a class of solutions if
$\varphi(x^{1}, x^{2})$ satisfies the Liouville equation with
coupling
constant proportional to $\alpha$. The most general solution of the
continual Toda equation has been constructed in the literature (see
[18] and
references therein), but it will not be needed for the present
purposes.
The solution
\be
e^{\Psi} = {{(x^{3})}^{2} - a^{2} \over 2 {\left(1 + {1 \over
8}({(x^{1})}^{2}
+ {(x^{2})}^{2})\right)}^{2}} ~, ~~~~ {(x^{3})}^{2} \geq a^{2}
\ee
provides an alternative description of the Eguchi--Hanson instanton
with
respect to its second (rotational) Killing symmetry.\footnote{The
Eguchi--Hanson
instanton has also a translational Killing symmetry in the sense that
the
corresponding $\Delta S_{-} = 0$. This can also be seen here from the
special
combination of the $x^{1}$ and $x^{2}$ coordinates that appear in eq.
(35).}
The solution
\be
\Psi = \log x^{3}
\ee
corresponds to the case
\be
{ds}^{2} = {(d x^{1})}^{2} + {(d x^{2})}^{2} + {1 \over x^{3}} {(d
x^{3})}^{2} +
x^{3} {(d \tau)}^{2} ~ ,
\ee
which gives rise to the trivial flat metric after performing the
coordinate
transformation
\be
w_{1} = 2 \sqrt{x^{3}} \cos(\tau /2) ~, ~~~~
w_{2} = 2 \sqrt{x^{3}} \sin(\tau /2) ~.
\ee

The important point that needs to be made for rotational Killing
symmetries is that
the T--S--T transformation is anomalous, thus breaking the
self--duality
of the original metric,
even though the resulting pure gravitational background is Ricci
flat. This is
a classical obstruction that has nothing to do with quantum
corrections.
The metric maintains its rotational symmetry by performing
the sequence of T--S--T transformations, but there is no coordinate
system of the
form (28), (29) for a new scalar function ${\Psi}^{\prime}$, as it
would have
been required by the theorem [13] for rotationally invariant
self--dual metrics.
We will see shortly that this problem, which can be observed in
Euclidean gravity
without ever employing the string viewpoint,
has a natural supersymmetric explanation in
the lowest order string effective theory.

The aparent incompatibility between S--duality and self--duality (and
hence
$N=4$ superconformal symmetry) can be easily verified by performing
the T--S--T
loop of operations
to the flat metric associated with the solution (36). Indeed, the
metric we
find at the end of the calculation  is
\be
{ds}^{2} = {x^{3} \over 1 - {(C x^{3})}^{2}} {\left( d \tau + C(x^{2}
d x^{1} -
x^{1} d x^{2})\right)}^{2} + {1- {(C x^{3})}^{2} \over x^{3}}
\left(x^{3} \left( {(d x^{1})}^{2} + {(d x^{2})}^{2} \right) +
{(d x^{3})}^{2} \right)
\ee
using for simpicity only elements of the Ehlers transform. It can be
verified
that for $C=0$ we obtain the original flat metric, while for any $C
\neq 0$
the resulting metric is not self--dual (nor anti self--dual).
Similarly, if we had applied the same method
to the Eguchi--Hanson instanton with respect to its rotational
symmetry,
a violation of self--duality would have also been observed.
Of course, as it was pointed
out earlier, S--duality would not lead to such a violation if the
T--S--T
transformation were performed using the translational Killing
symmetry of the
same solution. Therefore, on general grounds, if there is an anomaly
in
$\Delta S_{-}$, the S--duality appears to be incompatible with $N=4$
superconformal string theory.
Instead, the symmetry which seems compatible with self--duality in
this case
is $W_{\infty}$, at least in the pure gravitational sector of the
string
effective theory.

This result points to a problem which actually turns out to be
related
with the T--duality used to perform the combined T--S--T
transformation.
The reason is that for rotational Killing symmetries, unlike for the
translational ones, T--duality does not preserve the
(standard) space time supersymmetry.
The obstruction can be easily seen by considering the supersymmetric
variation
of the dilatino field $\lambda$. Remember that in the heterotic
string
context we are looking for bosonic ground states of a supersymmetric
theory.
The variation $\delta \lambda = 0$ implies in four dimensions that
the
dilaton and the anti--symmetric tensor fields have to satisfy
in the $\sigma$--model frame the consistency condition
\be
{1 \over 2} ~ H_{\mu \nu \rho} = \mp \sqrt{\det G} ~
{{\epsilon}_{\mu \nu \rho}}^{\sigma} {\partial}_{\sigma} \Phi
\ee
at each stage of the duality transformations. Since we have started
from
pure gravitational self--dual metrics, rather than anti self--dual,
the
appropriate sign in the equation above is minus. Taking into account
eq. (7),
it is straightforward to verify that the supersymmetric condition for
the dilatino field will be valid in the T--dual background only if
$S_{-}$
is constant and hence $\Delta S_{-} = 0$. Space time supersymmetry
can not
be preserved under T--duality for Killing vector fields with
$\Delta S_{-} > 0$. The subsequent application of the S and T
transformations
can only make things worse.

Supersymmetric T--duality transformations can be safely performed
only with
respect to translational Killing symmetries. We have already seen
that even
the simplest space time backgrounds, like the flat space and the
Eguchi--Hanson
instanton, have the potential for exhibiting an anomalous
supersymmetric
behaviour under T--duality when a rotational Killing symmetry is
used.
Solutions with more than one Killing symmetry are in practice much
easier
to construct.
It is worth emphasing, however, that solutions with  rotational
Killing
symmetries usually admit a translational Killing symmetry as well,
which
can be used safely. The reason is that if a space time admits two
Killing
symmetries, both of them rotational, it will not be a generic
real Euclidean solution
of the self--duality conditions [13]. The requirement of reality will
exclude this
possibility if the two rotational Killing vectors $K_{1}$ and $K_{2}$
are assumed
to form a closed algebra, i.e.,
\be
[K_{1} ~ , K_{2}] = \alpha K_{1} + \beta K_{2} ~ .
\ee
A way out is thus provided by the existence of an extra translational
Killing vector field leading to an $SU(2)$ symmetry algebra.
The multi--center Gibbons--Hawking instantons with $n > 2$ have only
one Killing
symmetry which is translational and hence safe under T--duality.

There are no real self--dual Euclidean solutions known to this date
which exhibit only one rotational Killing symmetry with no other
symmetries of either type.
New series
of well--behaved gravitational instatons can only be obtained by
investigating carefully the space time interpretation of the general
solution of the continual Toda equation [18].
If purely rotational instantons exist,
T--duality will never be compatible with their supersymmetry. The
problem
we are addressing here has also some relation with the non--abelian
duality transformations of gravitational string backgrounds [19]. The
presence of a rotational Killing vector field is a common element in
both cases. For this reason we think that space time supersymmetry
can not
be preserved by non--abelian duality. We hope to return to this
issue elsewhere taking into account gravitational backgrounds with
non--trivial axion and dilaton fields as well.

We note finally that if both S and T are symmetries of a certain
class
of 4--dim string
backgrounds with one Killing symmetry and with non--trivial
antisymmetric
tensor and dilaton fields, there will be two independent $SL(2, R)$
symmetries of the corresponding effective theory, namely S and
T--S--T. It
seems that in this way the combination of the S and T duality
transformations
enlarges the symmetry group of the theory to $O(2, 2)$.
This result is also expected from the reduced form of the 4--dim
effective action.
The reduction of pure gravity from four to three dimensions is
formulated as an $SL(2, R)/U(1)$ $\sigma$--model coupled to 3--dim
gravity, while the axion--dilaton system provides a second
$SL(2, R)/U(1)$ $\sigma$ model.
A similar proposal
was made recently for the compactification of the heterotic string
on a seven dimensional torus, where the T--duality group $O(7, 23)$
and
S--duality combine by intertwing into $O(8, 24)$ [20].
Going one dimension lower we expect the string duality group to
become an infinite dimensional
discrete subgroup of $\hat{O} (8, 24)$, in analogy with the infinite
dimensional Geroch group $\hat{O} (2, 2)$ of 4--dim strings with two
Killing symmetries [21]. The intertwining of the S and T
transformations
produces infinitely many new transformations in two dimensions.
The implications of such huge symmetries to the spectrum
of the eight dimensional
compactification of the heterotic string, are also under
consideration [22].

In summary, we found that $\Delta S_{-}$ is an index for the Killing
symmetry
that determines whether T--duality preserves supersymmetry. An
obstruction
was observed for $\Delta S_{-} \neq 0$. Its generalization to more
arbitrary
backgrounds is an interesting problem. We also gave a reformulation
of
the Ehlers--Geroch hidden symmetry of the reduced theory of 4--dim
gravity
in terms of the S--duality transformations of the lowest order string
effective theory. It was considered as providing a space time
interpretation of
S--duality. The simplest example was the transformation of the flat
space to
the Taub--NUT background.

\vskip 1cm
\centerline{\bf Acknowledgments}

\noindent
I am grateful to Amit Giveon and Elias Kiritsis for collaboration on
various aspects of the present work and for sharing their views on
the
problem of string duality symmetries. I have also benefited from
conversations with Louis Alvarez--Gaume, Massimo Bianchi, Sergio
Ferrara,
Bernard Julia, Costas Kounnas and Dieter Lust.

\vskip1.5cm
\centerline{\bf REFERENCES}
\begin{enumerate}
\item A. Font, L. Ibanez, D. Lust and F. Quevedo, Phys. Lett.
\underline{B249} (1990) 35.
\item J.H. Schwarz and A. Sen, Phys. Lett. \underline{B312} (1993)
105;
J.H. Schwarz, {\em ``Does String Theory Have a Duality Symmetry
Relating
Weak and Strong Coupling?"}, preprint CALT--68--1879,
hep--th/9307121,
July 1993;
A. Sen, {\em ``Strong--Weak Coupling Duality in Four Dimensional
String
Theory"}, preprint TIFR/TH/94--03, hep--th/9402002, February 1994.
\item J.W. van Holten, {\em ``Supersymmetry and the Geometry of
Taub--NUT"},
preprint NIKHEF--H/94--29, hep--th/9409139, September 1994.
\item A. Dabholkar, G. Gibbons, J.A. Harvey and F. Ruiz--Ruiz, Nucl.
Phys.
\underline{B340} (1990) 33; A. Strominger, Nucl. Phys.
\underline{B343} (1990) 167;
C. Callan, J.A. Harvey and A. Strominger, Nucl. Phys.
\underline{B359} (1991) 611,
\underline{B367} (1991) 60.
\item S.J. Rey, Phys. Rev. \underline{D43} (1991) 526;
M. Duff and J.X. Lu, Nucl. Phys. \underline{B354} (1991) 141;
R. Khuri, Nucl. Phys. \underline{B387} (1992) 315;
E. Kiritsis, C. Kounnas and D. Lust, Int. J. Mod. Phys.
\underline{A9} (1994) 1361.
\item M. Bianchi, F. Fucito, G.C. Rossi and M. Martellini, {\em ``ALE
Instantons
in String Effective Theory"}, preprint ROM2F--94--17,
hep--th/9409037,
September 1994.
\item I. Bakas and E. Kiritsis, work in progress.
\item E. Fradkin and A. Tseytlin, Nucl. Phys.
\underline{B261} (1985) 1;
C. Callan, D. Friedan, E. Martinec and M. Perry, Nucl. Phys.
\underline{B262} (1985) 593.
\item A. Giveon, M. Porrati and E. Rabinovici, {\em ``Target Space
Duality in
String Theory"}, preprint RI--1--94, hep--th/9401139, January 1994.
\item J. Ehlers, in {\em ``Les Theories Relativistes de la
Gravitation"},
CNRS, Paris (1959); R. Geroch, J. Math. Phys. \underline{12} (1971)
918.
\item G. Gibbons and S. Hawking, Commun. Math. Phys. \underline{66}
(1979) 291.
\item K.P. Tod and R.S. Ward, Proc. R. Soc. Lond. \underline{A368}
(1979) 411.
\item C.P. Boyer and J.D. Finley, J. Math. Phys. \underline{23}
(1982) 1126;
J.D. Gegenberg and A. Das, Gen. Rel. Grav. \underline{16} (1984) 817.
\item T. Eguchi, P.B. Gilkey and A.J. Hanson, Phys. Rep.
\underline{66} (1980) 213.
\item N. Hitchin, Math. Proc. Camb. Phil. Soc. \underline{85} (1979)
465;
P.B. Kronheimer, J. Diff. Geom. \underline{29} (1989) 665, 685.
\item A. Giveon and E. Kiritsis, Nucl. Phys. \underline{B411} (1994)
487;
E. Kiritsis and C. Kounnas, Phys. Lett. \underline{B331} (1994) 51.
\item I. Bakas, in the {\em ``Proceedings of the Trieste Conference
on
Supermembranes and Physics in 2+1 Dimensions"}, eds. M. Duff, C. Pope
and E. Sezgin,
World Scientific, Singapore (1990);
Q.H. Park, Phys. Lett. \underline{B236} (1990) 429; J. Avan, Phys.
Lett.
\underline{A168} (1992) 363; S.Y. Lou, Phys. Rev. Lett.
\underline{71} (1993) 4099.
\item M.V. Saveliev, Theor. Math. Phys. \underline{92} (1992) 457;
M.V. Saveliev and S.A. Savelieva, Phys. Lett. \underline{B313} (1993)
55.
\item X. de la Ossa and F. Quevedo, Nucl. Phys. \underline{B403}
(1993) 377;
A. Giveon and M. Rocek, Nucl. Phys. \underline{B421} (1994) 173;
E. Alvarez, L. Alvarez--Gaume and Y. Lozano, {\em ``On Non--Abelian
Duality"}, preprint CERN--TH.7204/94, hep--th/9403155, March 1994.
\item A. Sen, {\em ``Strong--Weak Coupling Duality in Three
Dimensional
String Theory"}, preprint TIFR/TH--94--19, hep--th/9408083, August
1994.
\item I. Bakas, {\em ``O(2,2) Transformations and the String Geroch
Group"},
preprint CERN--TH.7144/94, hep--th/9402016, February 1994.
\item I. Bakas and A. Giveon, work in progress.
\end{enumerate}

\end{document}